\documentstyle[11pt,paspconf,epsf]{article}

\begin{document}

\title{Molecular Clouds in Cooling Flow Clusters of Galaxies}
\footnote{Poster presented for the IGRAP International Conference 
{\it Clustering at high redshifts}, Marseille (France) June 29$^{th}$- July 
2nd, 1999 }
\author{L. Grenacher, Ph. Jetzer, D. Puy}
\affil{Paul Scherrer Institut, Laboratory for Astrophysics, Villigen
\\
and 
\\
Institute of Theoretical Physics, University of Zurich (Switzerland)}

\begin{abstract}
In many clusters of galaxies there is evidence for cooling flows in the central regions. 
A possibility is that a fraction of the gas forms cold molecular clouds. We investigate the 
minimum temperature which can be reached by clouds in cooling flows by computing the cooling function due to $H_2$, $HD$ and $CO$ molecules. As an example, we determine the minimum temperature achievable by clouds in the cooling flow 
of PKS 0745-191. 
\end{abstract}

\keywords{cluster of galaxies}

\section{Introduction}

Cooling flows in clusters of galaxies deposit large quantities of cool gas 
around the central galaxy, which is still growing. The final evolution of the cool gas is not clear. It may just accumulate as cool dense clouds.
The metallicity of a cluster seems to be correlated with the presence of a cooling flow. In this context molecules such as $CO$ can subsequently be formed in 
the gaseous medium. 
O'Dea, Baum, Maloney et al. (1994) searched for molecular gas, 
by looking for $CO$ emission lines in a heterogeneous sample of 
cooling flow clusters. They come to the conclusion 
that in order to have escaped detection the gas has to be very cold, close to 
the temperature of the Cosmic Background Radiation. 
\\
The aim of this contribution is to discuss the minimum temperature achievable
  by sub-clouds (resulting from the fragmentation of bigger clouds) in 
cooling flow regions by improving the analysis done by O'Dea, Baum, 
  Maloney et al. 
  (1994), in particular by considering clouds made of $H_2$, $HD$ and $CO$ 
  molecules and by computing cooling functions which are more appropriate for 
  temperatures below 20 K. 
\section{Thermal equilibrium}
In the standard Big Bang model primordial chemistry took place around the 
epoch of recombination. At this stage the chemical species were essentially 
hydrogen, deuterium, helium and lithium. Then with the adiabatic cooling of 
the Universe due to the expansion, different routes led to molecular formation 
(Puy et al. 1993). Molecules play an important role in the 
cooling of the denser clouds via the excitation of rotational levels. 
\\
In order to calculate analytically the molecular cooling, Puy, Grenacher and 
Jetzer (1999) consider only the transition between the ground state and 
the first rotational level for $H_2$, $HD$ and $CO$. As expected, it turns 
out that $CO$ is the 
main coolant at low temperatures below 20 K.
\\
The clouds are embedded in the hot intracluster gas, whose emission is 
dominated by thermal bremsstrahlung. We introduce an attenuating factor $\tau$ 
characterizing the column density surrounding the sub-clouds. The attenuated bremsstrahlung flux coming from the intracluster gas heats the clouds 
located in the cooling flow at a distance $r$ from the cluster center.
\\
Thermal balance between heating and cooling defines an equilibrium temperature 
of the sub-clouds at a distance $r$ inside the cooling region: $r<r_{cool}$ 
(where $r_{cool}$ is the cooling radius).
\\
As an example we choose PKS 0745-191 which is embedded in one of the largest 
known cooling flows. We adopt the following column densities 
for a typical small cloud: $N_{CO}=10^{14}$ cm$^{-2}$, 
$N_{H_2}=2 \times 10^{18}$ cm$^{-2}$ with $n_{H_2}=10^6$ cm$^{-3}$ (density of $H_2$). These values correspond to a size for the cloud of $L\sim 10^{-6}$ pc, moreover we take the 
following abundance $\eta_{HD}=N_{HD}/N_{H_2}=7 \times 10^{-5}$.
\\
The first column of the following table gives the minimum temperature of the 
sub-clouds, $T_{clump}$, inside the cooling flow region for different values 
of $\tau$ with $\eta_{CO}=5 \times 10^{-5}$, whereas the second column gives 
$T_{clump}$ for different values of $\eta_{CO}$ with $\tau=2.5$.

\begin{table}
\begin{flushleft}
\begin{tabular}{||l|l||l||l|l||}
\hline
\hline

{\bf $T_{clump}$} in K & $\tau$ & & {\bf $T_{clump}$} in K & $\eta_{CO}$
\\
with $\eta_{CO}=5 \times 10^{-5}$ & & & with $\tau=2.5$ & 
\\
\hline
\hline
{\bf 73} & 0.01 & & {\bf 13.5} & $10^{-5}$  
\\
\hline
{\bf 25} & 0.25 & & {\bf 3.5} & $2.5 \times 10^{-5}$ 
\\
\hline
{\bf 15} & 0.5 & & {\bf 3} & $5 \times 10^{-5}$ 
\\
\hline
{\bf 5} & 1 & & {\bf 3} & $7.5 \times 10^{-5}$ 
\\
\hline
{\bf 3.5} & 1.5 & & {\bf 3} & $10^{-4}$ 
\\
\hline
{\bf 3} & 2.5 & &  &  
\\
\hline
\hline
\end{tabular}
\end{flushleft}
\end{table}

The hypothesis of very cold molecular gas in cooling flows seems reasonable given also the fact 
that with our approximations we get upper limits for the cloud temperatures.

\acknowledgments 
We are grateful to Monique Signore for valuable discussions. This work has 
been supported by the {\it D$^r$ Tomalla Foundation} and by the Swiss National 
Science Foundation.

\end{document}